\documentclass[reprint, amsmath, amssymb, aps, floatfix, prl, superscriptaddress]{revtex4-2}

\usepackage{graphicx}
\usepackage{dcolumn}
\usepackage{bm}
\usepackage{orcidlink}
\usepackage{physics}
\usepackage{mhchem}
\usepackage{empheq}
\usepackage{siunitx} 
\usepackage{tcolorbox}
\newtcolorbox{mybox}[1]{fonttitle=\bfseries,title=#1}

\newcommand{\superR}{\hat{\hat{R}}}
\usepackage{appendix}

\begin{document}
\title{Interradical motion can push magnetosensing precision towards quantum limits}

\author{Luke D.\ Smith\orcidlink{0000-0002-6255-2252}}
\affiliation{Department of Physics and Astronomy, University of Exeter, Exeter, Devon, EX4 4QL, United Kingdom}
\affiliation{Living Systems Institute, University of Exeter, Stocker Road, Exeter, Devon, EX4 4QD, United Kingdom}

\author{Farhan T.\ Chowdhury\orcidlink{0000-0001-8229-2374}}
\affiliation{Department of Physics and Astronomy, University of Exeter, Exeter, Devon, EX4 4QL, United Kingdom}
\affiliation{Living Systems Institute, University of Exeter, Stocker Road, Exeter, Devon, EX4 4QD, United Kingdom}
\affiliation{
Department of Chemistry and Biochemistry, The Ohio State University, Columbus, Ohio 43210, United States}

\author{Jonas Glatthard\orcidlink{0000-0002-8411-4958}}
\affiliation{School of Physics and Astronomy, University of Nottingham, Nottingham NG7 2RD, United Kingdom}

\author{Daniel R.\ Kattnig\orcidlink{0000-0003-4236-2627}}
\affiliation{Department of Physics and Astronomy, University of Exeter, Exeter, Devon, EX4 4QL, United Kingdom}
\affiliation{Living Systems Institute, University of Exeter, Stocker Road, Exeter, Devon, EX4 4QD, United Kingdom}

\date{\today}

\begin{abstract}

Magnetosensitive spin-correlated radical-pairs (SCRPs) offer a promising platform for noise-robust quantum metrology. However, unavoidable interradical interactions, such as electron-electron dipolar and exchange couplings, alongside deleterious perturbations resulting from intrinsic radical motion, typically degrade their potential as magnetometers. In contrast to this, we show how structured molecular motion modulating interradical interactions in a \emph{live} chemical sensor in cryptochrome can, in fact, increase sensitivity and, more so, push precision in estimating magnetic field directions closer to the quantum Cram\'{e}r-Rao bound, suggesting near-optimal metrological performance. Remarkably, this approach to optimality is amplified under environmental noise and persists with increasing complexity of the spin system, suggesting that perturbations inherent to such natural systems have enabled them to operate closer to the quantum limit to more effectively extract information from the weak geomagnetic field. This insight opens the possibility of channeling the underlying physical principles of motion-induced modulation of electron spin-spin interactions towards devising efficient handles over emerging molecular quantum information technologies.

\end{abstract}
\maketitle

\textit{Introduction.}---The sensitivity of spin-correlated radical-pairs (SCRPs) to weak magnetic fields presents exciting avenues for potential applications across quantum technologies and biophysics \cite{hore25, liu25, Lin24_MolEng, mani22, yu21, kobori21, birad90}. In a physiological context, SCRPs in the protein cryptochrome are widely considered candidate magnetoreceptors of a chemical compass sense for migratory navigation \cite{Xu2021_Cry, Wiltschko2019_MagnetoBird, Hore2016_RPM, Ritz2000_Cry}. Such SCRPs function in noisy and complex condensed-phase regimes at ambient temperature while subjected to strong interradical interactions, such as electron-electron dipolar (EED) and exchange couplings, and environmentally induced structural fluctuations \cite{birad24, Gruning2024_SpinRelax, Babcock2020_EED, Kattnig2016_ElecSpinRelax}. In engineered biradicals \cite{birad20, birad19, katt18, birad12}, such effects usually suppress the coherent singlet-triplet interconversion that underpins their metrological sensitivity. This raises fundamental questions: How close to quantum limits could SCRPs operate under noisy biological conditions? And could nature have optimized SCRPs in the open \cite{Bartolke2021_SecCry, mei15, Liedvogel2010_Cry, Cashmore1999_Cry} to enable optimal magnetosensitivity?

Recent work comparing SCRP models in terms of the quantum Cram\'{e}r--Rao bound (QCRB) \cite{opt24, Holevo2011_QCRB}, which sets a fundamental limit on parameter estimation precision \cite{bayat25, levmet25}, showed a partial approach towards the QCRB in the limits of quantum system complexity, i.e.\ when increasing the number of hyperfine couplings towards biologically representative numbers. However, the precision fell short of that of the QCRB by more than an order of magnitude. In this work, we demonstrate that intrinsic interradical motion can drive magnetosensing in SCRPs to over 90\% of the quantum limit, achieving up to sub-degree angular precision, challenging the prevalent notion that environmental noise and molecular motion would preclude sensitivity near the quantum limit. Specifically, we consider driven radical-pair systems, motivated by prior findings showing how Landau-Zener-St\"{u}ckelberg-Majorana (LZSM) transitions \cite{Ivakhenko2023_LZSM} can enhance chemical signal contrast \cite{drive22}, with similarly periodic driving also having been shown to boost precision in quantum thermometry \cite{glatthard22}. We establish quantum limits by considering the relationship between radical motion and the QCRB for FAD$^{\bullet -}$/W$_{\mathrm{C}}^{\bullet+}$ as a representative radical-pair in cryptochrome, with our results informing more broadly about systems subject to electron spin-spin interactions, such as EED and/or exchange couplings.  

\begin{figure}[t]
\includegraphics[width=.9\linewidth]{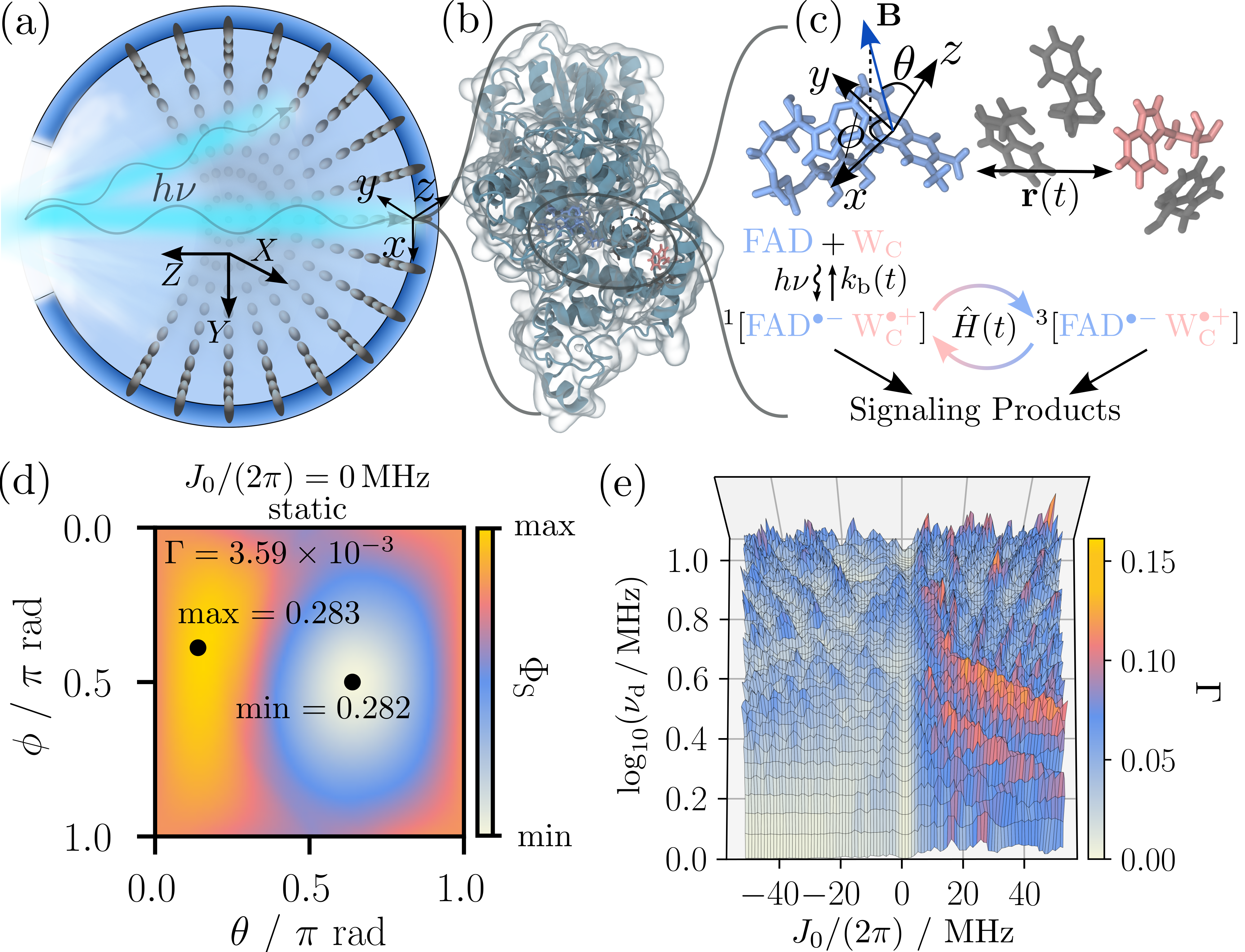}
\caption{\label{Fig:schematic} Magnetosensing modulated by interradical motion. (a) Schematic of a magnetosensitive array (in the eye-frame $(X, Y, Z)$), composed of magnetoreceptor units (gray spheroids; molecular frame $(x,y,z)$) distributed with fixed orientations along the retina. Incident blue light photoexcites radical-pairs within the magnetoreceptors. (b) Blue-light-activated magnetoreceptor protein cryptochrome. (c) Within cryptochrome, an FAD$^{\bullet -}$/W$_{\mathrm{C}}^{\bullet+}$ radical-pair is created in the singlet state ($^{1}[\cdot]$) and interconverts with triplet states ($^{3}[\cdot]$) via the interradical-separation-modulated Hamiltonian $\hat{H}(r(t))$, which includes Zeeman, hyperfine, exchange and electron-electron dipolar (EED) interactions. Spin-selective recombination occurs with rate $k_{\mathrm{b}}(t)$, while signaling products form at a rate $k_{\mathrm{f}}$. The Zeeman interaction imparts magnetosensitivity depending on the field orientation ($\theta$, $\phi$) in the molecular frame. (d) Angular dependence of the singlet yield $\Phi_{\mathrm{S}}$ and relative anisotropy $\Gamma = [\max(\Phi_{\mathrm{S}})-\min(\Phi_{\mathrm{S}})]/\overline{\Phi}_{\mathrm{S}}$ sampled over $37\times37$ magnetic field orientations ($\theta\ \& \ \phi \in [0, \pi]$) for a static FAD$^{\bullet -}$/W$_{\mathrm{C}}^{\bullet+}$ pair (fixed $\mathbf{r}(t)$) including EED interactions with exchange coupling $J_{0}/(2\pi)=0\,$MHz, and N$5$ and N$1$ hyperfine couplings on FAD$^{\bullet -}$ and W$_{\mathrm{C}}^{\bullet+}$, respectively. (e) Harmonic driving of the interradical separation at frequency $\nu_{\mathrm{d}}$ enhances the relative anisotropy, as shown for $48\times100$ combinations of $\nu_{\mathrm{d}}$ ($1\leq \nu_{\mathrm{d}}\leq 10$\,MHz) and $J_{0}$ ($-50\leq J_{0}/(2\pi) \leq 50$\,MHz) with EED interactions included.}
\end{figure}

\textit{Radical-Pair Model.}---We consider radical-pairs interacting with an external magnetic field and magnetic nuclei via Zeeman and hyperfine interactions, respectively, as shown in Fig.~\ref{Fig:schematic}(a)--(c). The time-independent part of the Hamiltonian defining these interactions is
\begin{align}
 \hat H_{0} &= {\hat H_{\rm{Zeeman}}} + {\hat H_{\rm{Hyperfine}}} \nonumber  \\
 &= \sum_{i}{\vec\omega _i} \cdot {{\mathbf{\hat S}}_i} + \sum_{i}\sum\limits_j {{{{\mathbf{\hat S}}}_i} \cdot {{\mathbf{A}}_{i,j}} \cdot {{{\mathbf{\hat I}}}_{i,j}}}, \label{eq:H0}
\end{align}
where the Larmor precession frequency $\vec{\omega}_{i} = - \gamma_{i}\mathbf{B}$ is given by the product of the magnetic field $\mathbf{B} = B_{0}(\sin \theta \cos \phi, \sin \theta \sin \phi, \cos \theta)$, with $B_{0}$ denoting the magnitude ($\sim50\,\mu$T for the geomagnetic field), and the electron gyromagnetic ratios. The electron spin in the $i$th radical is denoted ${\mathbf{\hat S}}_{i}$, whilst the nuclear spin ${\mathbf{\hat I}}_{i,j}$ is coupled to the $i$th electron spin via hyperfine coupling tensors ${{\mathbf{A}}_{i,j}}$. Additionally, we introduce time-dependent dipolar and exchange interactions between electron spins and random-field relaxation (RFR). The time-dependent part of the Hamiltonian that defines these interactions is
\begin{align}
    \hat{H}_{1}(t) =& \hat{H}_{\mathrm{Dipolar}}(t) + \hat{H}_{\mathrm{Exchange}}(t) + \hat{H}_{\mathrm{RFR}}(t) \nonumber \\
    =&  \hat{\mathbf{S}}_1 \cdot \mathbf{D} (t) \cdot  \hat{\mathbf{S}}_2  - 2J(t) \hat{\mathbf{S}}_1 \cdot \hat{\mathbf{S}}_2  \nonumber\\ &+ \sum_{i = 1,2} \sum _{j = x,y,z} \mathbf{b}_{i,j}(t) \cdot \hat{\mathbf{S}}_{i,j},\label{eq:H1}
\end{align}
where $\mathbf{D} (t)$ is the electron-electron dipolar (EED) coupling tensor, $J(t)$ is the exchange coupling, and $\mathbf{b}_{i,j}(t)$ denotes uncorrelated magnetic field noise of equal strength across the three Cartesian coordinates. The EED tensor and exchange coupling are proportional to  $r^{-3}(t)$, and $\exp(-\beta r(t))$ with $\beta=1.4$\,\AA$^{-1}$ \cite{Moser92}, respectively. We consider harmonically driven \cite{drive22} and optimally controlled \cite{cont24, drive22} interradical modulations. The harmonic driving interradical modulation takes the form $r(t) = \frac{\Delta_{\mathrm{d}}}{2}[1-\cos(2\pi \nu_{\mathrm{d}}t)] + r_0$, where $\nu_{\mathrm{d}}$ is the driving frequency and the driving amplitude is set to $\Delta_{\mathrm{d}} = 3\,$\AA, which was previously found to produce significant sensitivity enhancements while constituting a relatively modest motion \cite{drive22}. The interradical distance of the static FAD$^{\bullet -}$/W$_{\mathrm{C}}^{\bullet+}$ radical-pair is specified with $r_0=17.2\,$\AA\, informed from molecular dynamics simulations \cite{Gruning2022_Dynamical}. Further details on the form of the dipolar coupling and how the optimally controlled modulations are obtained  are provided in the Appendix, with parameters reported in the Supplemental Material (SM). 

Following the photoinduced electron transfer reaction that forms the candidate magnetoreceptor FAD$^{\bullet -}$/W$_{\mathrm{C}}^{\bullet+}$, the initial state is an electronic singlet state, given by a spin density operator $\hat{\rho}(0) = \frac{\hat{P}_{\mathrm{S}}}{Z},$ where $\hat{P}_{\mathrm{S}}$ is the singlet projection operator and $Z=Z_{\mathrm{A}}Z_{\mathrm{B}}$ denotes the dimension of the nuclear subspace associated with the two radicals. This initial state evolves under the total Hamiltonian $\hat{H}(t)=\hat{H}_{0} + \hat{H}_{1}(t)$ via the master equation $\partial_{t}{\hat{\rho}_{\theta}}(t) = -i[\hat{H}(t), \hat{\rho}_{\theta}(t)] - \hat{\hat{K}}(t)\hat{\rho}_{\theta}(t)$,
where, $\hat{\hat{K}}(t)$ is the reaction operator comprising the reaction rate for product formation $k_{\mathrm{f}}$ and time-dependent recombination $k_{\mathrm{b}}(t)$, $[\cdot, \cdot]$ indicates the commutator, and $\hat{\rho}_{\theta} = \hat{\rho}(t, \theta, \phi)$ signifies that the state acquires sensitivity to the magnetic field inclination through the Zeeman interaction. In this work, we restrict to $k_{\mathrm{b}}(0)=k_{\mathrm{f}}=1\mu$s$^{-1}$, reflecting the asymmetry in the reaction channels from singlet and triplet states typically found in cryptochromes \cite{Hore2016_RPM, Maeda2012_Asymm_CRY}.

From $\hat{\rho}_{\theta}(t)$, the singlet population can be calculated as $p_{\mathrm{S}}(t) = \mathrm{Tr}[\hat{P}_{\mathrm{S}}\hat{\rho}_{\theta}(t)]$, with the
singlet yield defined as 
\begin{align}
    \Phi_{\mathrm{S}} = \int_{0}^{\infty} k_{\mathrm{b}}(t)p_{\mathrm{S}}(t)\,\mathrm{d}t.
\end{align}
From this the established relative anisotropy measure of sensitivity that quantifies the contrast in recombination yields over all magnetic-field directions is obtained by
$\Gamma = (\Phi_{\mathrm{S}, \mathrm{max}} - \Phi_{\mathrm{S}, \mathrm{min}})/\overline{\Phi}_{\mathrm{S}}$
where $\Phi_{\mathrm{S}, \mathrm{max}}$ and $\Phi_{\mathrm{S}, \mathrm{min}}$ represent the maximum and minimum recombination yields and $\overline{\Phi}_{\mathrm{S}}$ indicates the mean recombination yield. Fig.~\ref{Fig:schematic}(d)\&(e) demonstrate that the modulation of interradical motion can increase yield contrast and $\Gamma$, and has been considered in further depth in \cite{drive22}. 

\textit{Quantum Metrology.}---Following the framework established in \cite{opt24}, we define the probe state as the steady state $\hat{\rho}_{\mathrm{ss}}$ of continuous generation and recombination, evaluated from $\hat{\rho}_{\theta}(t)$ with specific details provided in the Appendix. As the electron spins couple to the magnetic field via the Zeeman interaction, the electronic part of $\hat{\rho}_{\mathrm{ss}}$ encodes the intensity and directional dependence of the magnetic field. The quantum Fisher information (QFI) \cite{Braunstein1994} and the associated quantum Cram\'er--Rao bound (QCRB) \cite{Holevo2011_QCRB} constrain the maximal precision attainable in the statistical estimation of a parameter $\vartheta$ from measurements on a quantum state $\hat{\rho}_\vartheta$. For an unbiased estimator based on $N$ independent measurements, the bound is provided by $\frac{1}{\delta^2\vartheta}\leq N F_\vartheta\leq N\mathcal{F}_\vartheta$ where $\delta^2\vartheta$ represents the variance of the $\vartheta$-estimates, $F(\vartheta)$ is the classical Fisher information (CFI) \cite{Fisher1925, Paris2011} for any observable and $\mathcal{F}(\vartheta)$ is the QFI. Specifically, the classical Fisher information establishes the sensitivity of the state to a particular set of measurements and takes the form, 
\begin{align}\label{eq:defcfi}
F_\vartheta &= \sum\nolimits_n p_{n,\vartheta}\, (\partial_\vartheta \ln p_{n,\vartheta})^2 
= \sum_{n} \frac{(\partial_{\vartheta}p_{n,\vartheta})^{2}}{p_{n,\vartheta}},
\end{align}
where $p_{n,\vartheta} = \mathrm{Tr}[\hat{\rho}_\vartheta \hat{\Pi}_{n}]$ are the Born rule probabilities associated with the measurement element $\hat{\Pi}_{n}$, with $\sum_{n} \hat{\Pi}_{n} = \hat{I}$ and $\hat{I}$ denoting the identity operator, given that the state of the system before the measurement was $\hat{\rho}_\vartheta$. The QFI can be obtained by optimizing the CFI over all possible measurements. This framework can thus be used to analyze the quantum limit of precision in estimating a magnetic field parameter $\vartheta = \theta$ as realizable in SCRPs, by choosing a set of measurements for the CFI $\Pi_{n} \in \{ \hat{P}_{\mathrm{S}}, \hat{P}_{\mathrm{T}} \}$, where $\hat{P}_{\mathrm{S}}$ and $\hat{P}_{\mathrm{T}}=\hat{I}-\hat{P}_{\mathrm{S}}$, are the singlet and triplet projection operators, respectively. This represents measurements in the magnetosensitive basis associated with chemically distinguishable outcomes, for which spin-selective recombination reactions of the radical-pair is the only anticipated mechanism in natural systems. While both the CFI and singlet yield arise from the same singlet-triplet measurement channel, the yield constitutes a time-integrated, coarse-grained observable derived from the underlying spin-state probabilities. In contrast, the CFI, as defined in the Appendix, is expressed in terms of $\Theta_{\mathrm{S}}=\mathrm{Tr}[\hat{P}_{\mathrm{S}}\hat{\sigma}_{\mathrm{ss}}]$, where $\hat{\sigma}_{\mathrm{ss}}$ is the normalized steady-state spin density operator, quantifies the information accessible through the $\{\hat{P}_{\mathrm{S}}, \hat{P}_{\mathrm{T}}\}$ measurement channel applied to the steady-state spin ensemble. Equality is only recovered in the static-rate limit, for which $\Phi_{\mathrm{S}} \propto k_{\mathrm{b}}\Theta_{\mathrm{S}}$ and the yield carries the same $\theta$-dependence.  

Specifically, the ratio $F_{\theta}/\mathcal{F}_{\theta}$ can be used to assess the metrological efficiency of information extraction of $\theta$ encoded in the quantum state from singlet-triplet-based measurements in a SCRP, where $F_{\theta}/\mathcal{F}_{\theta}\leq1$ represents the degree of approach to QCRB and a value of $1$ represents its saturation. Here, we have chosen to report the maximum of the ratio as a representative metric, since this point will always be present and measurable within the magnetosensitive array (Fig.~\ref{Fig:schematic}(a)) of at least one eye, regardless of bird orientation. Moreover, because the precise biological mechanism for magnetosensory information processing remains unknown, averaging the ratio over all orientations may obscure the intrinsic sensory precision by including magnetoreceptors not contributing to the directional estimate. In contrast, the maximum indicates the configuration in which the system most effectively exploits its available information from singlet and triplet measurements, a quantity of interest both for biological comparison and for the design of artificial radical-pair magnetometers. Nonetheless, other metrics, such as orientation-averaged ratios, are also valid and could be more appropriate for other applications. In the SI, we therefore report the orientation-averaged ratios; qualitatively similar trends arise under under interradical motion for this alternative metric. 

While the ratio $F_{\theta}/\mathcal{F}_{\theta}$ quantifies the optimality of the measurement basis relative to the quantum limit, it does not by itself ensure large absolute estimation precision. For this reason, we also report relative anisotropy values, Fisher-information magnitudes, and the corresponding angular precision $\Delta \theta=1/\sqrt{NF_{\theta}}$ estimated for magnetoreceptor numbers ($N=2\times10^{5}$--$N=2\times10^{6}$) based on the energy resolution limit \cite{Kominis2025_EnergyResolutionAnimal} with realistic radical-pair lifetimes in the range $1$--$10\,\mu$s \cite{luo24, Kattnig2016_ElecSpinRelax, kattnig2016njp}. Although these receptor numbers have not yet been experimentally verified, they remain biologically plausible and conservative when compared with established sensory systems such as vision, where each photoreceptor cell contains approximately $10^{8}$ rhodopsins \cite{Milo2015_CellBioNumbers}. Our estimates are also consistent with information-theoretic analyses of magnetoreception based on photon flux and reaction yields \cite{Hiscock2019_Navigating_Night}. Taken together, these comparisons demonstrate that the observed approaches towards the QCRB can correspond to practically and biologically significant precision levels, accompanied by substantial chemical contrast. 
\begin{figure*}
\includegraphics[width=\linewidth]{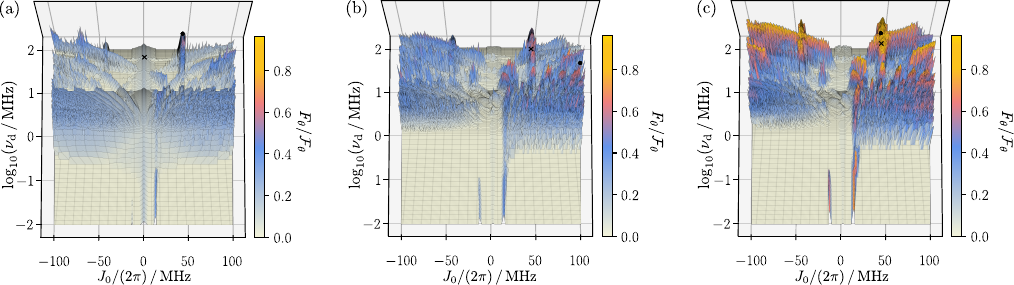}
\caption{\label{Fig:N5} Approach to the quantum Cram\'{e}r-Rao bound (QCRB) under driven dynamics and increasing interaction complexity in a simple radical-pair model comprising the N$5$ nucleus, representing the dominant hyperfine coupling of FAD$^{\bullet -}$/W$_{\mathrm{C}}^{\bullet+}$. The approach is assessed via the ratio of classical to quantum Fisher information $F_{\theta}/\mathcal{F}_{\theta}$, plotted as a function of $209\times200$ combinations of driving frequencies $\nu_{\mathrm{d}}$  and exchange couplings $J_{0}$ spanning $0.01\leq \nu_{\mathrm{d}}\leq 100$\,MHz and $-100\leq J_{0}/(2\pi) \leq 100$\,MHz, respectively. Each point corresponds to the maximum ratio achieved over 180 magnetic field orientations ($\theta \in [0, \pi]$ at fixed $\phi=0$). Panels show scenarios of increasing system complexity: (a) exchange only; (b) inclusion of electron-electron dipolar coupling (EED) interactions; (c) inclusion of both EED, and random-field noise (RFR). Circles indicate closest approach to the QCRB; crosses indicate highest chemical contrast $\Gamma$. For (a) the maximum ratio $78.3\%$ occurs at $J_{0}/(2\pi)  = 41\,$MHz, $\nu_{\mathrm{d}} = 47\,$MHz,  with $NF_{\theta} = 0.216$ (angular precision $\Delta \theta =2.76^{\circ}\times10^{-2}$--$8.72^{\circ}\times10^{-2}$). Incorporating EED and RFR broadens the region of high ratios, reaching $96.4\%$, while maximum $\Gamma$ values ($0.456$, $0.342$, and $0.197$) also occur at higher QCRB saturation ($12\%$, $56.1\%$, and $78\%$). Across all these cases, absolute Fisher-information magnitudes remain significant ($NF_{\theta}\geq  0.011$), corresponding to angular precisions better than $\Delta \theta =1.23^{\circ}$. Closer approaches to the QCRB with EED and RFR thus occur within practical levels of precision and reflect more efficient extraction of the available information.}
\end{figure*}

\textit{Noise and interradical modulation.}---We evaluate the approach to the QCRB for a simple driven radical-pair model comprising nucleus N$5$, representing the dominant hyperfine coupling of FAD$^{\bullet -}$/W$_{\mathrm{C}}^{\bullet+}$. Fig. \ref{Fig:N5} shows the ratio of $F_{\theta}/\mathcal{F}_{\theta}$, which indicates how closely the system approaches the QCRB, for driving frequency $\nu_{\mathrm{d}}$ against exchange coupling $J_{0}$. Due to the approximate axial symmetry of the hyperfine interaction for this system \cite{opt24}, we scan over $\theta$ and set $\phi=0$. To evaluate how noise and interradical interactions \cite{nikuni25, hong23} influence metrological performance for the driven model, we consider three progressively complex scenarios in Fig.~\ref{Fig:N5}: (a) exchange only, neglecting EED interactions and relaxation, (b) inclusion of EED, and (c) inclusion of both EED and RFR (with relaxation rate $\gamma = 1\,\mu$s$^{-1}$). To maintain focus on regions of increased sensitivity, in panels (b) and (c), we restrict $F_{\theta}/\mathcal{F}_{\theta}$ to points where the relative anisotropy $\Gamma$ is maintained or improved relative to the undriven, exchange-interaction-free reference $\Gamma(J_{0}=0, \nu_{\mathrm{d}}=0)$. In panel (a), where the idealized absence of dipolar and relaxation effects leads to unrealistically high baseline sensitivities, we instead restrict $F_{\theta}/\mathcal{F}_{\theta}$ to points where the relative anisotropy does not fall below $10\%$ of the reference. This reflects the idea that a measurement, even if optimal, must induce a noticeable chemical change.

Interestingly, under moderate driving ($1\leq\nu_{d}\leq100$ MHz) within the range of biological plausibility as predicted from molecular dynamics simulations \cite{Benjamin2025_DynamicRadical}, the system can attain $96.4$\% saturation of the QCRB. Moreover, with the inclusion of EED and RFR, the optimization is generally more significant, especially in the frequency range of $0\leq \nu_{\mathrm{d}}\leq 10$\,MHz, which was previously found to support relative anisotropy enhancements \cite{drive22, katt14}. Although absolute magnitudes inevitably decrease with increasing environmental complexity, as expected for magnetoreceptive biological systems which operate at degree level accuracy \cite{Schwarze2016_Compass_5degree, Akesson2001_Avian_Orientation},  they remain substantial. Specifically, the maximal QFI reaches $N\mathcal{F_{\theta}}=32.4,\ 5.40,\  \& \ 0.153$ ($\Delta\theta = 7.12^{\circ} \times 10^{-3}$--$2.25^{\circ} \times 10^{-2}$, $1.74^{\circ} \times 10^{-2}$--$5.51^{\circ} \times 10^{-2}$, $0.104^{\circ}$--$0.323^{\circ}$), and the maximal CFI values reach $NF_{\theta}=1.22,\  0.708, \ \& \ 0.129$ ($\Delta\theta = 3.67^{\circ} \times 10^{-2}$--$0.116^{\circ}$, $4.81^{\circ} \times 10^{-2}$--$0.152^{\circ}$, $0.113^{\circ}$--$0.357^{\circ}$), for the exchange-only, with-EED, and with-RFR-and-EED scenarios, respectively. At these maximal values, the static system is vastly outperformed, with QFI smaller by two to three orders of magnitude, and CFI smaller by two to six orders of magnitude. This shows that driven interradical modulations not only optimize the system in approach to the QCRB but significantly increase the absolute performance by enhancing the magnitude of precision as compared to the static scenario. Additional simulations provided in the SM, including full plots of QFI and CFI magnitudes, relative anisotropy, and orientation-averaged values confirm these trends, and show that even field-orientation averaged values attain ratios of up to $78.2\%$ with mean $NF_{\theta}=2.41\times10^{-2}$ ($\Delta \theta = 0.261^{\circ}$--$0.825^{\circ}$). 

\begin{figure*}
\includegraphics[width=0.85\linewidth]{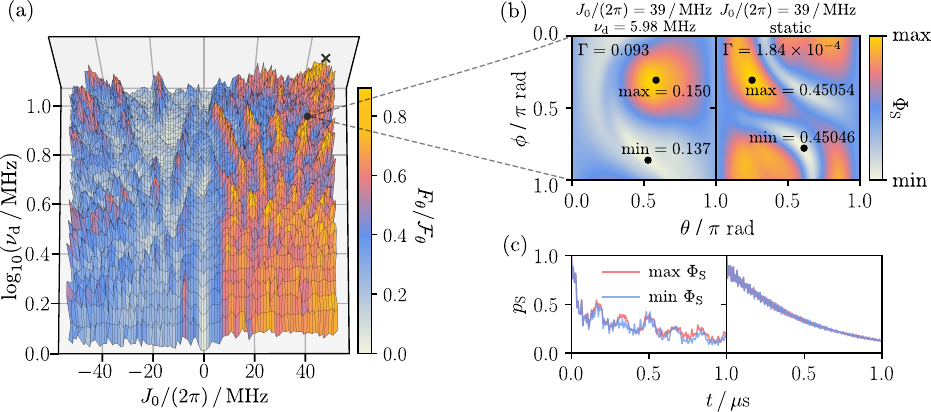}
\caption{Approach to the quantum Cram\'{e}r-Rao bound (QCRB) under driven dynamics in a representative radical-pair model of FAD$^{\bullet -}$/W$_{\mathrm{C}}^{\bullet+}$ comprising N$5$ and N$1$ hyperfine couplings, electron-electron dipolar and exchange interactions. (a) The QCRB is assessed via the ratio of classical to quantum Fisher information $F_{\theta}/\mathcal{F}_{\theta}$, plotted as a function of $48\times100$ combinations of driving frequencies $\nu_{\mathrm{d}}$ and exchange couplings $J_{0}$ spanning $1\leq \nu_{\mathrm{d}}\leq 10$\,MHz and $-50\leq J_{0}/(2\pi) \leq 50$\,MHz, respectively. Each point corresponds to the maximum value achieved over $37\times37$ magnetic field orientations ($\theta\ \& \ \phi \in [0, \pi]$). The closest approach to the QCRB is indicated by a circle while the cross indicates highest chemical contrast $\Gamma=0.161$ with $86.8\%$ QCRB saturation and $NF_{\theta}=0.161$ (angular precision $\Delta\theta=0.101^{\circ}$--$0.319^{\circ}$).(b) The singlet yield $\Phi_{\mathrm{S}}$ profile for the closest approach to the QCRB of $F_{\theta}/\mathcal{F}_{\theta} = 89.5\%$, which occurs at $J_{0}/(2\pi)  = 39\,$MHz, $\nu_{\mathrm{d}} = 5.98\,$MHz, with $NF_{\theta}=0.015$ ($\Delta\theta=0.327^{\circ}$--$1.04^{\circ}$) is shown and compared to the static system, with relative anisotropy $\Gamma$ indicated. (c) Singlet probability $p_{\mathrm{S}}$ over time that gives rise to the maximum and minimum singlet yields. Overall, driving provides close approaches to the QCRB, providing a larger contrast in yield and better defined scanning axis, originating from a restoration of coherent oscillations in the singlet probability at the driving frequency.\label{Fig:N5N1} }
\end{figure*}

\textit{Biologically motivated system.}---Moving toward a more biologically representative model of FAD$^{\bullet -}$/W$_{\mathrm{C}}^{\bullet+}$, we include the N$5$ and N${1}$ hyperfine couplings, each localized on a different radical, with EED and exchange interactions included. Fig.~\ref{Fig:N5N1}(a) shows the ratio of $F_{\theta}/\mathcal{F}_{\theta}$ to indicate the approach to the QCRB, for driving frequency $\nu_{\mathrm{d}}$ against exchange coupling $J_{0}$ within the range identified as beneficial in the N$5$-system. As axial symmetry is broken in this model system, we sampled over $\theta \ \& \ \phi$. $F_{\theta}/\mathcal{F_{\theta}}$ is displayed at all points, for which there is an improvement over the static reference in the magnitudes of relative anisotropy, CFI, and QFI. We evaluated all quantities over orientations to the magnetic field $\theta$ and $\phi$ and took the maximum as representative, as a particular scanning axis through varying reaction yields could be selected or have evolved to contain the maximal point, while qualitatively similar results emerge for the average, as we show in the SM. In the Appendix, we provide further details on how this scheme can be viewed as a measurement of the total electronic spin, with the CFI equivalent to the error in measuring the square of the total angular momentum $\hat{S}^{2}$, the relation to reaction yields, and evaluation of the QFI.  

Close approaches to the QCRB are observed over a broad range of exchange coupling and driving frequencies. At the optimal point, where $F_{\theta}/\mathcal{F}_{\theta} = 89.5\%$ with $NF_{\theta}=0.015$ ($\Delta\theta=0.327^{\circ}$--$1.04^{\circ}$), the singlet yield $\Phi_{\mathrm{S}}$ profile is plotted in Fig.~\ref{Fig:N5N1}(b) alongside its static counterpart. Driving significantly enhances both yield contrast and the relative anisotropy, increasing $\Gamma$ from $1.84\times10^{-4}$ in the static system to $0.093$ under driving. Unlike the static case, where yield fluctuations obscure a scanning axis through $\theta$ and $\phi$, the driven yield profile exhibits a clear high-yield and low-yield point, establishing a well-defined scanning axis. Taking the points of maximum and minimum $\Phi_{\mathrm{S}}$ and plotting the associated singlet probabilities $p_{\mathrm{S}}$ in Fig.~\ref{Fig:N5N1}(c), shows that the enhancement is due to the restoration of coherent singlet-triplet oscillations, driven at the modulation frequency. Although driving can generally break the field-inversion symmetry, we find the effect to be minor: at the closest approach to the QCRB the relative anisotropy differs by only $0.065\%$ and the average yield by $1.46\times10^{-4}\%$, when comparing one hemisphere with the other, as shown in the SM. We note also that the degree of approach to the QCRB from the inclusion of driving is broadly maintained on increasing the number of hyperfine couplings from one to two. Additional simulations are provided in the SM, including full plots of QFI and CFI magnitudes, relative anisotropy, and orientation-averaged values confirm these trends, which show that orientation averaged values broadly attain ratios up to $34\%$ with mean $NF_{\theta}=6.16\times10^{-3}$ ($\Delta \theta = 0.516^{\circ}\times10^{-3}$--$1.63^{\circ}$). 

\begin{figure*}
\includegraphics[width=\linewidth]{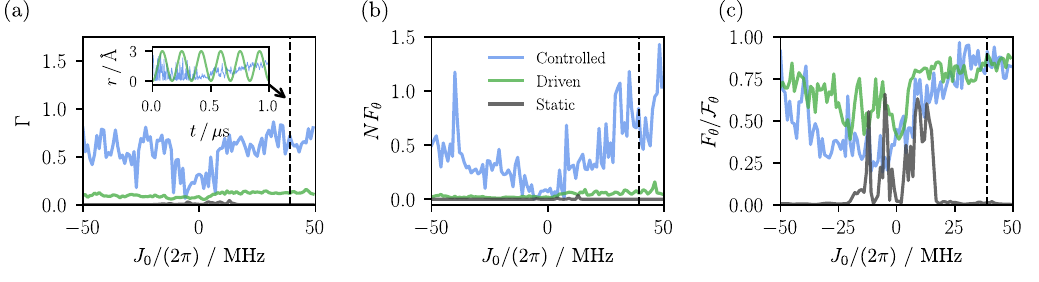}
\caption{\label{Fig:control} Comparison of quantum-controlled interradical fluctuations that maximize yield contrast, harmonically driven fluctuations, and the static system. (a) Relative anisotropy ($\Gamma$), with inset showing the modulated interradical distance for both control and driven cases at the optimal exchange coupling value $J_{0}/(2\pi)=39\,$MHz. (b) the maximal Classical Fisher information ($NF_{\theta}$), and (c) the maximal ratios ($F_{\theta}/\mathcal{F_{\theta}}$), calculated over $37\times 37$ combinations of $\theta \ \& \ \phi \in [0, \pi]$ across the exchange coupling range $-50\leq J_{0}/(2\pi) \leq 50\,\mathrm{MHz}$. Quantum-controlled fluctuations generate the strongest enhancements, followed by the driven system, both of which significantly outperform the static system.}
\end{figure*}

\textit{Precision under quantum control.}---Thus far, harmonic modulations of recombination, exchange and electron-electron dipolar interactions have been shown to enhance magnetic field sensing through improving yield contrast, precision, and the approach to quantum limits as assessed by the CFI and the QCRB. To assess whether more complex interradical modulation can show further improvement, we leverage optimal quantum control \cite{wolf25, oqc24, cont24, woods24, llobet20, cont15, qc10}, conditioned upon maximizing the yield contrast ($\Phi_{\mathrm{S, max}}-\Phi_{\mathrm{S,min}}$) of the biologically representative FAD$^{\bullet -}$/W$_{\mathrm{C}}^{\bullet+}$ model, to derive optimized interradical fluctuations. This is compared to the static and harmonically driven system in Fig.~\ref{Fig:control} by generating (a) the relative anisotropy ($\Gamma$), (b) the CFI ($NF_{\theta}$), (c) and $F_{\theta}/\mathcal{F_{\theta}}$ plotted against exchange coupling.  
Control results exhibit a marked enhancement across majority of $J_{0}$ values, for relative anisotropy, CFI, and QFI, reaching as high as $\Gamma = 0.863$, $NF_{\theta}=1.43$ ($\Delta \theta = 3.39^{\circ}\times10^{-2}$--$0.107^{\circ}$), and $N\mathcal{F}_{\theta}=3.68$ ($\Delta \theta = 2.11^{\circ}\times10^{-2}$--$6.68^{\circ}\times10^{-2}$). The harmonically driven system is generally lower, with maximum values at $\Gamma = 0.161$, $NF_{\theta}=0.161$ ($\Delta \theta = 0.101^{\circ}$--$0.319^{\circ}$), and $N\mathcal{F}_{\theta}=0.503$ ($\Delta \theta = 5.71^{\circ}\times 10^{-2}$--$0.181^{\circ}$), yet it still marks a significant improvement over the maximum values of the static system at $\Gamma = 0.044$, $NF_{\theta}=0.036$ ($\Delta \theta = 0.213^{\circ}$--$0.674^{\circ}$), and $N\mathcal{F}_{\theta}=0.064$ ($\Delta \theta = 0.160^{\circ}$--$0.507^{\circ}$). Moreover, the static system for the representative scenario of $J_{0}/(2\pi)=0\,$MHz, falls to as low as $9.11^{\circ}$ -- $28.8^{\circ}$, corresponding to $NF_{\theta}= 1.98 \times10^{-5}$. With control, approaches to the QCRB as close as $F_{\theta}/\mathcal{F}_{\theta} = 97\%$ can be reached.

\begin{figure}[t]
\includegraphics[width=.9\linewidth]{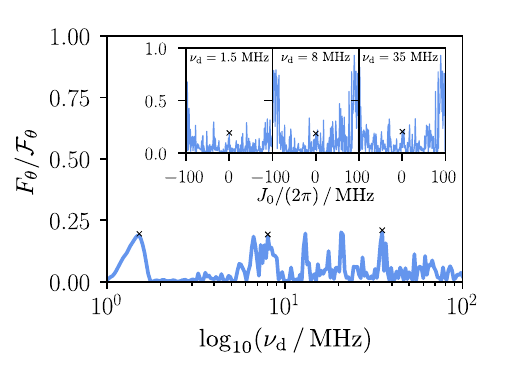}
\caption{\label{Fig:robustness} Robustness of approach to the quantum Cram\'{e}r-Rao bound (QCRB) under driven dynamics of FAD$^{\bullet -}$/W$_{\mathrm{C}}^{\bullet+}$ with four hyperfine couplings (N$5$ and N$10$ in FAD; N$1$ and H$1$ in Trp) electron-electron dipolar and exchange interactions. Approach is assessed by the maximal $F_{\theta}/\mathcal{F}_{\theta}$ for the case of $J_{0}/(2\pi)=0\,$MHz with $200$ driving frequencies in the range $1\leq \nu_{\mathrm{d}} \leq 100\,$MHz, while insets extend the exchange coupling to $200$ values in the range $-100\leq J_{0}/(2\pi) \leq 100\,$MHz for select driving frequencies $\nu_{\mathrm{d}}=1.5,\ 8, \ \& \ 35\,$MHz. The maximal $F_{\theta}/\mathcal{F}_{\theta}$ is evaluated for orientations maximizing and minimizing yield difference in the static case. This shows that approaches to the QCRB are maintained on increase of hyperfine complexity.}
\end{figure}

\textit{Robustness to biophysical complexity.}---
While relative anisotropy, and therefore chemical contrast, is known to decrease as the number of hyperfine-coupled nuclei increases in FAD$^{\bullet -}$/W$_{\mathrm{C}}^{\bullet+}$, the factor of enhancement from interradical motion persists \cite{drive22}. To test the robustness of the approach to the QCRB under increasing hyperfine complexity, we extend the model to include four hyperfine couplings (two per radical), adding the next most relevant hyperfine couplings N$10$ and H$1$. Fig.~\ref{Fig:robustness} shows $F_{\theta}/\mathcal{F}_{\theta}$, over $200$ driving frequencies in the range $1\leq \nu_{\mathrm{d}} \leq 100\,$MHz in the presence of EED interactions with $J_{0}/(2\pi) = 0\,$MHz, where insets show selected driving values of $\nu_{\mathrm{d}}=1.5,\ 8, \ \& \ 35\,$MHz where the ratio peaks, plotted over the range $-100\leq J_{0}/(2\pi) \leq 100\,$MHz. For computational tractability, only the orientations of $\theta$ and $\phi$ that maximize and minimize the static-system yield difference are scanned. Despite this minimal sampling, near-saturation of the QCRB is reached (up to $F_{\theta}/\mathcal{F}_{\theta}=93\%$), suggesting that the approach to the QCRB from interradical fluctuations is maintained as hyperfine couplings are added. To quantify the impact on absolute precision, we show that the four hyperfine-coupled system exhibits a maximum $NF_{\theta}= 0.247$ ($\Delta \theta = 8.15^{\circ}\times10^{-2}$--$0.258^{\circ}$), maintaining substantial precision. Extrapolating to biologically realistic regimes with tens of hyperfine couplings \cite{qTr25}, suggests that harmonically driven systems could yield angular precisions on the order of degrees or better. Although RFR noise would reduce the CFI by roughly an order of magnitude, structured modulations such as those identified through quantum-control \cite{cont25}, could compensate for this loss. In contrast, static systems of comparable complexity perform far worse: previous analyses \cite{opt24} report $NF_{\theta} \sim 1\times 10^{-7}$ even without RFR, corresponding to $\Delta\theta = 128^{\circ}$ -- $405^{\circ}$. These results demonstrate that both optimality and practical precision enhancements arising from interradical motion are preserved as hyperfine interaction complexity increases.  

\begin{figure*}
\includegraphics[width=\linewidth]{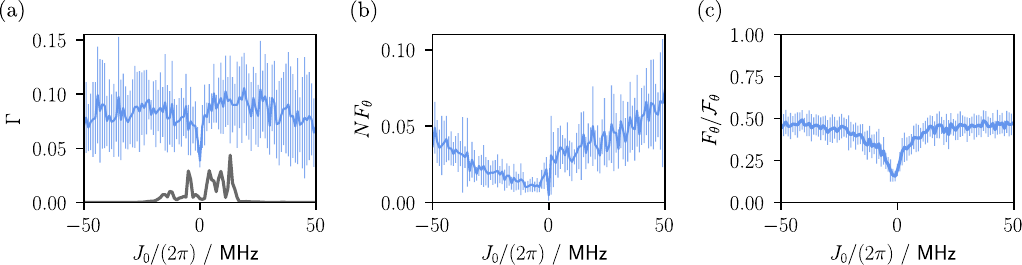}
\caption{Mean and standard error over ten phase trajectories of a biophysically inspired interradical modulation informed by molecular dynamics, shown for (a) relative anisotropy ($\Gamma$), with static-system comparison, (b) classical Fisher information ($NF_{\theta}$), and (c) the optimality ratio ($F_{\theta}/\mathcal{F}_{\theta}$). For each trajectory, values are calculated over $37\times 37$ combinations of $\theta \ \& \ \phi \in [0, \pi]$ across the exchange coupling range $-50\leq J_{0}/(2\pi) \leq 50\,\mathrm{MHz}$, with maxima reported. The system attains up to $F_{\theta}/\mathcal{F}_{\theta}\approx50\%$, $NF_{\theta} = 0.0741$ ($\Delta \theta = 0.149^{\circ}$--$0.471^{\circ}$), and $\Gamma\approx0.10$, demonstrating that interradical motion continues to enhance precision, approach to the quantum Cram\'{e}r-Rao bound, and chemical contrast, under realistic, composite modulations.} \label{Fig:nat_mod}
\end{figure*}

As we assume monochromatic modulations so far, it is apt considering if more biophysically realistic dynamics can yield comparable effects. To explore this, we use estimates of dominant frequencies of motion $\omega_{i}$ and their relative amplitudes $c_{i}$ from molecular dynamics simulations \cite{Benjamin2025_DynamicRadical}, constructing a composite modulation (up to a random phase) scaled to $\Delta_{\mathrm{d}}=3\,$\AA: $r(t) = \Delta_{\mathrm{d}}[\sum_{i} c_{i} \cos(\omega_{i}t + \phi_{i})] + r_{0}$, with $\phi_{i}\in[0, 2\pi]$.  These modulations explore shorter interradical distances and thus stronger couplings. Fig.~\ref{Fig:nat_mod} shows mean and standard error over ten phase trajectories in the model with two hyperfine couplings for (a) $\Gamma$, (b) $NF_{\theta}$, and (c) $F_{\theta}/\mathcal{F_{\theta}}$, against $J_{0}$. The results confirm that the effect broadly persists under realistic composite modulations: $\Gamma$ attains $\approx0.10$ ($0.05\lesssim \Gamma \lesssim 0.15$), $F_{\theta}/\mathcal{F_{\theta}}$ reaches approximately $50\%$, and maximal CFI reaches $NF_{\theta} = 0.0741$ ($\Delta \theta = 0.149^{\circ}$--$0.471^{\circ}$), while maximal QFI is $N\mathcal{F}_{\theta} = 0.533$  ($\Delta \theta = 5.55^{\circ}\times10^{-2}$--$0.175^{\circ}$). This shows that, despite extending to smaller interradical distances, realistic fluctuations can retain much of the metrological enhancement identified under harmonic and controlled modulations. A path toward even greater biomimetic performance may lie in selectively inhibiting shorter range motion and reinforcing modulation frequencies conducive to enhanced magnetosensory precision

\textit{Discussion.}---Experiments \textit{in vitro} suggest a photo-induced FAD$^{\bullet -}$/W$_{\mathrm{C}}^{\bullet+}$ radical-pair \cite{Ramsay2024_CoarseMD, Frederiksen2023_MutationalTrp, Wong2021_CryFourTrp, Kattnig2016_ChemAmpMFE, Nohr2016_ElectronTransfer, Muller2015_DiscFourthTrp, Giovani2003_LightinducedCry}, generated via electron transfer from a tryptophan residue ($\mathrm{W_{C}}$) to a flavin adenine dinucleotide (FAD) cofactor, as a potential SCRP underpinning a biochemical compass sensor. An alternative candidate, FADH$^{\bullet}$/O$^{\bullet-}_2$, formed in the dark during flavin reoxidation with molecular oxygen, is also occasionally deliberated \cite{Deviers2024_CryBindSuperoxide, Salerno2023_LongTimeSuperoxide, Arthaut2017_LightIndReactOx, Vanwilderen2015_KineticStudiesOxidation, Muller2011_LightCrySuperoxide, Ritz2009_MagCompMol, Massey1994_ActMolOx}. Both share the dilemma that, while simple models of their spin dynamics suggest their feasibility to underpin a compass in principle, models of realistic complexity accounting for the effects of quantum noise predict only marginal sensitivity, casting doubts on their ability to provide an effective sensory modality. To resolve this, additional enhancement mechanisms have been proposed, including exchange/dipolar cancellation \cite{Efimova08}, three-radical mechanisms \cite{Kattnig17a, Keens18}, radical scavenging \cite{Kattnig17}, relaxation-assisted mechanisms \cite{luo24, kattnig2016njp}, diffusive \cite{Ramsay23} and driven-radical motion \cite{Benjamin2025_DynamicRadical, drive22, kobori21}, the quantum Zeno effect \cite{binhi25, Denton2024_Zeno, Bugarth2020_ZenoGenOp, Dellis2012}, and chirality-induced spin selectivity \cite{Foo2025, Smith2025_CISSZeno, Tiwari2022_RoleCISS, Luo2021_CISSRad}. Various metrics, such as energy resolution limits \cite{Kominis2025_EnergyResolutionAnimal}, mutual information \cite{Ren2021_Angular_Precision, Hiscock2019_Navigating_Night}, Wigner-Yanase skew information \cite{Kominis2023_WignerYanase}, and the QFI \cite{opt24, Kominis2017_MagnetometersQFI, Guo2017_QFI}, have been explored, relating the sensing performance to the number of magnetoreceptors, coherence, and measurement strategy.

Nevertheless, most studies of enhancement mechanisms focus on the chemical reaction yield contrast over magnetic field orientation, which does not directly establish whether the ultimate limits of quantum precision are approached, nor identify how high the ceiling of magnetometric performance goes. Our approach, on the other hand, provides more direct insight into how nature may have harnessed structural adaptations from biophysics to leverage environment-assisted quantum dynamics \cite{noise25, bec25, noise24, Dodin_2021}, principles of which could be exploited in tunable molecular quantum technologies \cite{Lin24_MolEng, mani22, yu21} for information processing and metrology. We expect our results to generalize beyond FAD$^{\bullet -}$/W$_{\mathrm{C}}^{\bullet+}$ to radical systems in which modulations can arise naturally or can be engineered, with their underlying physical principles potentially also offering a route towards identifying biomimetic design principles for quantum sensing. Particularly promising are systems with reference-probe characteristics \cite{Lee2014_AltRadCry, Cai2012_QuantLimHyp, Ritz2010_Optimal, Timmel2001_ModelCalcHyp}, such as $\mathrm{FADH}^{\bullet}/ \mathrm{O_{2}}^{\bullet -}$ or the hypothetical $\mathrm{FAD}^{\bullet -}/ \mathrm{Z}^{\bullet}$ \cite{Lee2014_AltRadCry}. Future work could further investigate manipulation of structured modulations informed from molecular dynamics simulations, where periodic motion with frequency components in the beneficial $1$--$10\,$MHz range have indeed been recently observed \cite{Benjamin2025_DynamicRadical}, and explore whether alternative mechanisms, such as the quantum Zeno effect in tightly bound radicals \cite{Denton2024_Zeno}, can similarly approach performance near quantum limits. 

\textit{Conclusion.}---We show that interradical modulations induced by molecular motion can push the angular precision in magnetic field sensing towards the limit set by the quantum Cram\'{e}r-Rao bound, by increasing the metrological efficiency, quantified through the ratio $F_{\theta}/\mathcal{F_{\theta}}$, while simultaneously enhancing the absolute magnitudes of precision. Remarkably, realistic system complexities, including exchange and dipolar couplings, random field relaxation noise, and increased hyperfine couplings, were found to augment metrological efficiency. This originates from restored singlet-triplet coherence, driven by Landau-Zener-St\"{u}ckelberg-Majorana transitions \cite{Ivakhenko2023_LZSM, drive22, segal14} that increase the chemical yield contrast and both classical and quantum Fisher information magnitudes. While even harmonic modulations are effective, modulations identified through quantum control \cite{rwcont25, maeda25, martino24bang, cont24}, and combining harmonic components at $1$--$10\,$MHz with components resonant with hyperfine splittings, enable substantially greater precision. Such modulations, which may have emerged in cryptochromes over their $541$--$1000$ million years of evolution from DNA photolyases \cite{Bartolke2021_SecCry, mei15, Liedvogel2010_Cry, Cashmore1999_Cry}, reduce the untenable angular errors predicted for the static system (exceeding $100^{\circ}$) to within the biological functional window of degrees \cite{cai24, Hiscock2019_Navigating_Night, Schwarze2016_Compass_5degree, Akesson2001_Avian_Orientation}, and offer scope for engineering sub-degree improvements beyond what nature may have achieved.

\begin{acknowledgments}
\textit{Acknowledgements.}---The authors acknowledge support from the Office of Naval Research Global (ONR-G Award Number N62909-21-1-2018), the Engineering and Physical Sciences Research Council (EPSRC grants EP/V047175/1 and EP/X027376/1) and the Biotechnology and Biological Sciences Research Council (BBSRC grants BB/Y514147/1, BB/Y51312X/1). JG is supported by a Leverhulme Trust Research Project Grant (RPG-2023-177). The authors further acknowledge the use of the University of Exeter High Performance Computing Facility. For the purpose of open access, the authors have applied a Creative Commons Attribution (CC BY) license to any Author Accepted Manuscript version arising from this submission.
\end{acknowledgments}

\setcounter{equation}{0}
\setcounter{figure}{0}
\renewcommand{\theequation}{A\arabic{equation}}
\renewcommand{\thefigure}{A\arabic{figure}}

\clearpage

\begin{appendices}

\section{Appendix}

\textit{Harmonically driven interradical motion.}---
Following the approach in \cite{drive22}, we introduce radical motion
\begin{align}
r(t) = \frac{\Delta_{\mathrm{d}}}{2}[1-\cos(2\pi \nu_{\mathrm{d}}t)] + r_0,
\end{align}
for which the corresponding radical motion modulated dipolar coupling is
\begin{align}
        \hat{H}_{\rm Dipolar}(r(t)) =& - d(r(t))\Big[3(\hat{\mathbf{S}}_1 \cdot \mathbf{v})(\hat{\mathbf{S}}_2\cdot \mathbf{v}) - \hat{\mathbf{S}}_1 \cdot \hat{\mathbf{S}}_2\Big],
    \label{eq:eed_hamiltonian}
\end{align}
where $d(r(t)) \equiv \mu_0g_e^2\mu_B^2/4\pi r(t)^3 > 0$, with $r\equiv|\mathbf{r}|$ and constants taking their usual definitions, and $\mathbf{v}=\mathbf{r}/r$. The recombination \cite{drive22} is modulated as
\begin{align}
k_{\mathrm{b}}(r(t)) = k_{\mathrm{b_{0}}} \exp[-\beta (r(t)-r_0)].
\end{align}
The exchange interaction is modulated in the same functional form as
\begin{align}
    \hat{H}_{\mathrm{Exchange}}(r(t)) = -2 J(r(t)) \hat{\mathbf{S}}_1 \cdot \hat{\mathbf{S}}_2, \label{eq:exchange_hamiltonian}
\end{align}
with $J(r(t)) = J_{0} \exp[-\beta (r(t)-r_0)]$. Together with Eqs.~\ref{eq:H0} and \ref{eq:H1} these define  
\begin{align}
    \hat{H}_{\mathrm{eff}}(r(t)) = \hat{H}(r(t)) - i\left( \frac{k_{b}(r(t))}{2}\hat{P}_{\mathrm{S}} + \frac{k_{f}}{2}\hat{I}\right). \label{eq:eff_hamiltonian}
\end{align}
Using above and following \cite{opt24}, we define the probe state as the radical-pair in a steady-state of continuous generation and recombination as $\hat{\rho}_{\mathrm{ss}} = \frac{k_{0}c}{Z} \Tilde{G}(s=0)\hat{P}_{\mathrm{S}}$, with $k_{0}$ denoting the generation rate, $c$ the (constant) concentration of cryptochrome in the resting state, and $\Tilde{G}(s)$ is the Laplace transform of $\hat{\hat{G}}(t)$.

\textit{Explicit CFI and QFI formulations.}---
Extending the results of \cite{opt24} to a system under time-dependent interradical motion, the CFI using Eq.~\ref{eq:defcfi} with measurement elements $\hat{\Pi}_{n} \in \{ \hat{P}_{\mathrm{S}}, \hat{P}_{\mathrm{T}} \}$ takes the form
\begin{align}
    F_{\theta} &= \frac{(\partial_{\theta}\mathrm{Tr}[\hat{P}_{\mathrm{S}}\hat{\sigma}_{\mathrm{ss}}])^{2}}{\mathrm{Tr}[\hat{P}_{\mathrm{S}}\hat{\sigma}_{\mathrm{ss}}]} 
    + \frac{(\partial_{\theta}\mathrm{Tr}[(\hat{I} - \hat{P}_{\mathrm{S}})\hat{\sigma}_{\mathrm{ss}}])^{2}}{\mathrm{Tr}[(\hat{I} - \hat{P}_{\mathrm{S}})\hat{\sigma}_{\mathrm{ss}}]} \nonumber \\
    &= \frac{(\partial_{\theta} \Theta_{\mathrm{S}})^{2}}{\Theta_{\mathrm{S}}(1 - \Theta_{\mathrm{S}})}
\end{align}
where $\hat{\sigma}_{\mathrm{ss}}$ is the normalized spin density operator and $\Theta_{\mathrm{S}} = \mathrm{Tr}[\hat{P}_{\mathrm{S}}\hat{\sigma}_{\mathrm{ss}}] $ represents the the expectation value of the conditional singlet probability, which is the probability of finding the system in the singlet state, conditioned on its survival. 
Following the derivation in \cite{opt24}, it follows that $1/\Delta^{2}\theta = NF_{\theta}$ still holds when considering the error associated with measuring observable $\hat{O}=\hat{S}^{2}$ through $N$ independent measurements. However, we note that $\Theta_{\mathrm{S}}$ differs from the singlet yield $\Phi_{\mathrm{S}}$ due to the time-dependent nature of the recombination rate constant, and the relation to singlet yield is only reclaimed in the limit of the static system (up to normalization).

To obtain the QFI \cite{Paris2011}, we note that $\partial_{\vartheta}p_{n,\vartheta} = \mathrm{Tr}[\partial_{\vartheta} \hat{\rho}_\vartheta \hat{\Pi}_{n}] = \mathrm{Tr}[\hat{\rho}_\vartheta \hat{\Pi}_{n}\hat{L}_{\vartheta}]$, where $\hat{L}_{\vartheta}$ is the symmetric logarithmic derivative. Maximizing the CFI over $\{ \hat{\Pi}_{n} \}$ gives $\mathrm{Tr} \left( \hat{L}_{\vartheta}^2 \hat{\rho}_\vartheta \right)= \mathcal{F}_\vartheta$. In the basis that diagonalizes $\hat{\rho}_\vartheta$, and using the spectral decomposition ${\hat{\rho}_\vartheta = \sum_{i} p_{i} \vert \psi_{i} \rangle \langle \psi_{i} \vert}$, we obtain
\begin{align}
    \mathcal{F}_\vartheta = 2 \sum_{p_{i}+p_{j} \neq 0} \frac{1}{p_{i} + p_{j}} \big\vert \langle \psi_{i} \vert \partial_{\vartheta}\hat{\rho}_{\vartheta} \vert \psi_{j} \rangle \big \vert^{2}.
\end{align}
Alternative approaches to compute the QFI are possible if $\hat{\rho}_\vartheta$ is invertible \cite{safranek18}. 

\textit{Optimal quantum control.}--- We discretize our control functions into piecewise-constant segments, as standard in gradient-based approaches \cite{dey25, oqc05}. The total Hamiltonian including the drift and control parts, with interradical displacement $r_j$ constant during each interval, is
\begin{equation}\label{eq:discrete_hamiltonian}
\hat{H}_{\mathrm{Total}}(r_j) = \hat{H}_{0} + \hat{H}_{\mathrm{Exchange}}(r_{j}) + \hat{H}_{\mathrm{Dipolar}}(r_{j}),
\end{equation}
where the interradical displacement is related to the control modulation $u_{j}$ as $r_{j}=\Delta_{\mathrm{max}}u_{j}+r_{0}$ with $0\le u_{j} \le 1$. Following the approach from \cite{cont24}, the controls are optimized to maximize the difference in yields: $\Phi_{\mathrm{S}, \mathrm{max}} - \Phi_{\mathrm{S}, \mathrm{min}}$, with min and max in this case referring to the magnetic field orientations of the minimal and maximal yield respectively in the absence of the controlled distance fluctuations. To numerically solve the maximization problem, we use 1000 controls of $1\mskip3mu$ns each, whereby the maximal displacement is bound to $\Delta_{\mathrm{max}} = 3\,\mathrm{\AA}$, ensuring physical feasibility of controls. 

 \begin{figure}[t]
\includegraphics[width=\linewidth]{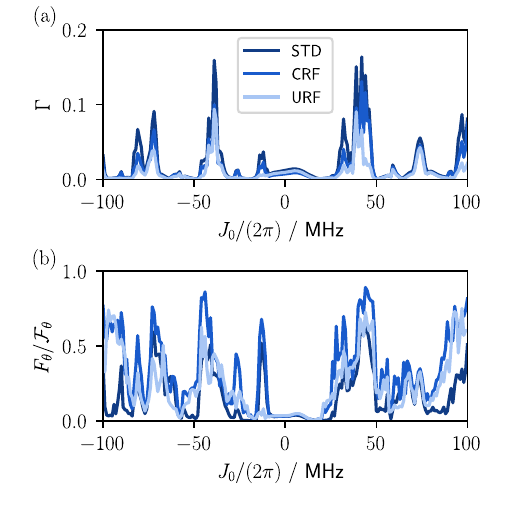}
\caption{Comparison of correlated random field (CRF), uncorrelated random field (URF), and singlet-triplet dephasing (STD) noise models under driven dynamics in a simple radical-pair model comprising the N$5$ nucleus, representing the dominant hyperfine coupling of FAD$^{\bullet -}$/W$_{\mathrm{C}}^{\bullet+}$. (a) relative anisotropy ($\Gamma$). (b) approach to the quantum Cram\'{e}r-Rao bound quantified via the maximal $F_{\theta}/\mathcal{F}_{\theta}$. Each data point has been calculated over 180 magnetic field orientations ($\theta \in [0, \pi]$ at fixed $\phi=0$) for a driving frequency $\nu_{\mathrm{d}}=35\,$MHz, which maximizes $\Gamma$ in the reference noise model of Fig.~\ref{Fig:N5}. \label{Fig:noise_models}.}
\end{figure}

 \textit{Additional noise models.}---
While Fig.~\ref{Fig:N5} examines the influcence of random field relaxation noise on the approach to the quantum Cram\'{e}r-Rao bound, and on the magnitudes of relative anisotropy and precision, chosen primarily as an indicator of the effects of noise, it is instructive to consider whether the same trends persist under alternative noise mechanisms. We thus compare three noise models: correlated random field (CRF), uncorrelated random field (URF), and singlet-triplet dephasing (STD). Their corresponding relaxation superoperators are given by
\begin{align}
    \superR_{\mathrm{CRF}}[\cdot] =& \gamma \sum_{\alpha=x,y,z}\mathcal{D}[\hat{S}_{1,\alpha}+\hat{S}_{2,\alpha}] \,\cdot,\\
        \superR_{\mathrm{URF}} [\cdot] =&  \gamma\sum_{i=1,2}\sum_{\alpha=x,y,z} \mathcal{D}[\hat{S}_{i,\alpha}] \,\cdot, \\
    \superR_{\mathrm{STD}}[\cdot] =& \gamma \mathcal{D}[\hat{P}_{\mathrm{S}}] \, \cdot ,
\end{align}
where $i$ labels the radical, $x,y,z$ denotes Cartesian components, and $\mathcal{D}$ represents the Lindblad dissipator, $\mathcal{D}[\hat{L}]\hat{\rho} = \hat{L}\hat{\rho}\hat{L}^{\dagger} - \frac{1}{2}\{\hat{L}^{\dagger}\hat{L}, \hat\rho\}$ with $\{\cdot, \cdot\}$ denoting the anticommutator. The uniform noise rate is set to $1\,\mu$s$^{-1}$. Fig.~\ref{Fig:noise_models} shows $\Gamma$ and $F_{\theta}/\mathcal{F}_{\theta}$ against $J_{0}$, for $\nu_{\mathrm{d}}=35\,$MHz, the driving frequency maximizing $\Gamma$ in the random field model of Fig.~\ref{Fig:N5}. The results indicate that qualitatively similar behavior arise across all three noise models, confirming that the observations are not specific to the particular noise mechanism considered. 

\begin{figure}[t]
\includegraphics[width=\linewidth]{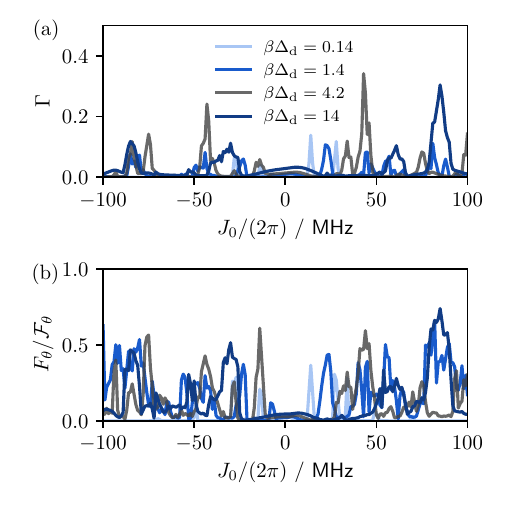}
\caption{Variation of the product $\beta\Delta_{\mathrm{d}}$ governing the effective modulation strength of the exchange coupling against $J_{0}$, in a simple radical-pair model comprising the N$5$ nucleus, representing the dominant hyperfine coupling of FAD$^{\bullet -}$/W$_{\mathrm{C}}^{\bullet+}$, and including electron-electron dipolar (EED) interactions. (a) relative anisotropy ($\Gamma$). (b) approach to the quantum Cram\'{e}r-Rao bound quantified via the maximal $F_{\theta}/\mathcal{F}_{\theta}$. Each data point has been calculated over 180 magnetic field orientations ($\theta \in [0, \pi]$ at fixed $\phi=0$) for a driving frequency $\nu_{\mathrm{d}}=36\,$MHz, which maximizes $\Gamma$ at the reference coupling $\beta\Delta_{\mathrm{d}}=3.4$, indicated in gray, with comparative results shown across the range $\beta\Delta_{\mathrm{d}}=0.14$--$14$ in blue corresponding to $\Delta_{\mathrm{d}}=0.1$--$10$\AA. \label{Fig:eff_amp_models}}
\end{figure}

\textit{Variation of modulation parameters.}---
Throughout the main part of this study, we have employed a fixed set of parameters for the modulation, such as an amplitude of $\Delta_{\mathrm{d}}=3$\AA\ and a decay constant of $\beta=1.4\,$\AA$^{-1}$, as is typical for proteins \cite{Moser92}. These choices were motivated by biological plausibility and prior studies \cite{drive22}, which also indicate that the effects of driven radical motion persist across a broad range of amplitudes. Nevertheless, it is of interest to explore variations, which is realised here in terms of the product $\Delta_{\mathrm{d}} \beta$, which quantifies the combined effects of modulation strength and interaction decay for the exchange interaction, while the dipolar interaction is modulated by $\Delta_{\mathrm{d}}$. Fig.~\ref{Fig:eff_amp_models} shows $\Gamma$ and $F_{\theta}/\mathcal{F}_{\theta}$ against $J_{0}$, for $\nu_{\mathrm{d}}=36\,$MHz, the driving frequency maximizing $\Gamma$ for the FAD$^{\bullet -}$/W$_{\mathrm{C}}^{\bullet+}$ pair comprising the N$5$ nucleus and including EED interactions. The results indicate that significant effects persist across $\beta\Delta_{\mathrm{d}}=1.4$--$14$, with similar peak values occurring at different $J_{0}$, suggesting these can be tuned alongside modulation frequencies.

\end{appendices}
\end{document}